\documentclass[pra,aps,superscriptaddress,twocolumn]{revtex4-1}

\usepackage{color}
\usepackage{graphicx}
\usepackage[utf8x]{inputenc}
\usepackage[T1]{fontenc}
\usepackage{siunitx}
\usepackage{amstext}
\usepackage{textgreek}
\usepackage{hyperref}
\usepackage{epstopdf}
\usepackage{amsfonts}
\usepackage{amsmath}
\usepackage{multirow}
\usepackage{dsfont}
\usepackage{physics}
\usepackage{bbold}
\usepackage{filecontents}
\usepackage{subfigure}
\graphicspath{{../},{figures/}}

\def\id{{\rm 1\kern-.22em l}}

\usepackage{marvosym}
\usepackage[rgb]{xcolor}

\makeatletter

\newcommand{\Rnum}[1]{\expandafter\@slowromancap\romannumeral #1@}

\usepackage[normalem]{ulem}

\begin{document}

\title{Measuring concurrence in qubit Werner states without aligned reference frame}

\author{Kateřina Jiráková} \email{katerina.jirakova@upol.cz}
\affiliation{Joint Laboratory of Optics of Palacký University and Institute of Physics of Czech Academy of Sciences, 17. listopadu 12, 771 46 Olomouc, Czech Republic}

\author{Artur Barasiński} \email{artur.barasinski@upol.cz}
\affiliation{Joint Laboratory of Optics of Palacký University and Institute of Physics of Czech Academy of Sciences, 17. listopadu 12, 771 46 Olomouc, Czech Republic}
\affiliation{Faculty of Physics, University of Wroc\l{}aw, plac Maxa Borna 9, PL-50-204 Wroc\l{}aw, Poland}

\author{Antonín Černoch} \email{antonin.cernoch@upol.cz}
\affiliation{Joint Laboratory of Optics of Palacký University and Institute of Physics of Czech Academy of Sciences, 17. listopadu 12, 771 46 Olomouc, Czech Republic}

\author{Karel Lemr}
\email{k.lemr@upol.cz}
\affiliation{Joint Laboratory of Optics of Palacký University and Institute of Physics of Czech Academy of Sciences, 17. listopadu 12, 771 46 Olomouc, Czech Republic}   

\author{Jan Soubusta} \email{soubusta@fzu.cz}
\affiliation{Institute of Physics of the Czech Academy of Sciences, Joint Laboratory of Optics of PU and IP AS CR, 17. listopadu 50A, 772 07 Olomouc, Czech Republic}

\begin{abstract}
The genuine concurrence is a standard quantifier of multipartite entanglement, detection and quantification of which still remains a difficult problem from both theoretical and experimental point of view. Although many efforts have been devoted toward the detection of multipartite entanglement (e.g., using entanglement witnesses), measuring the degree of multipartite entanglement, in general, requires some knowledge about an exact shape of a density matrix of the quantum state. An experimental reconstruction of such density matrix can be done by full state tomography which amounts to having the distant parties share a common reference frame and well calibrated devices. Although this assumption is typically made implicitly in theoretical works, establishing a common reference frame, as well as aligning and calibrating measurement devices in experimental situations are never trivial tasks. It is therefore an interesting and important question whether the requirements of having a shared reference frame and calibrated devices can be relaxed. In this work we study both theoretically and experimentally the genuine concurrence for the generalized Greenberger-Horne-Zeilinger states under randomly chosen measurements on a single qubits without a shared frame of reference and calibrated devices. We present the relation between genuine concurrence and so-called nonlocal volume, a recently introduced indicator of nonlocality.
\end{abstract}

\date{\today}

\maketitle

\section{Introduction}
\label{Introduction}


Secure and reliable information exchange is of paramount importance worldwide, hence practical implementation of quantum communications protocols outside the scientific laboratory has become one of the main focus of recent studies \cite{Sunnatphot10_2016,Valivarthinatphot10_2016}. Naturally, such advances in quantum communication methods require the ability to perform quantum measurements in an unstable environment, where the strict requirements for alignment and calibration of remote devices are hard to be met (e.g. a long-distance quantum communication \cite{Ursin2006NatPhys,Yinnature488_2012,Manature489_2012} or satellite-based communications \cite{Villoresi2008NJourPhys,Rennature549_2017,Yin2020Nat,Vallone2015PhysRevLett}). Specifically, the above-mentioned quantum communications experiments usually rely on quantum optical devices, where qubits are encoded into polarization states of light. However, this necessarily requires to share a common reference measurement frame that has to be well aligned and measurement devices calibrated (in a sense of well defined scale of measurement apparatus such as rotation angles of wave plates). Furthermore, it needs to be maintained stable for the entire experiment or communication as well. From an experimental point of view, this is, however, never achieved without technical difficulties.
Maintaining a common reference frame seems a trivial assumption when confined to a laboratory. But the long-distance quantum communications beyond Earth's surface \cite{Villoresi2008NJourPhys,Rennature549_2017,Yin2020Nat,Vallone2015PhysRevLett} has already led scientists to  re-evaluate the practicality of such assumption~\cite{Bonato2006OptExp,Han2020OptExp}. 

A possible solution of these problems in free space could be to use rotationally invariant states of light \cite{D'Ambrosio2012NatCom}. However, to the best of our knowledge, no one has yet applied neither of these solutions in satellite quantum communication. Instead, much attention has been paid to a so-called reference-frame independent (RFI) protocols \cite{Li2019OptLett,Tannous2019AppPhysLett,Souza2008PhysRevA,Laing2010PhysRevA,Chen2006PhysRevLett,Rezazadeh2010QuantInfProc,Liu2019PhysRevApplied,Guo2019EurLet}. 
For instance, it was proved in Ref. \cite{Yoon2019OptCom} that RFI quantum key distribution protocol \cite{Xue2020IntJTheoPhys} is more robust under reference frame fluctuations than its standard counterpart \cite{Bennett1984inproc,Bruss1998PhysRevLett}.

Motivated by all these observations, in this paper we also investigate the RFI approach. In particular, we focus on quantum entanglement which is undoubtedly an essence of many quantum information procedures \cite{Horodecki2009RevModPhys,Barasinskisr8_15209_2018,Barasinskiprl122_2019}. Therefore, it is necessary to be able to test the presence of entanglement and, for reason explained above, it is practical to manage it in RFI mode \cite{Wang2016SciBul,Lawson2014PhysRevA,Yang2020PhysRevA,Shadbolt2012SciRep, Wallman2012PhysRevA}.  
Over time several methods for entanglement detection under these constraints have been proposed. They are based on various approaches, for example, on the violation of a Bell inequality  \cite{Liangprl104_2010,Barasinskipra101_2020}, the second moment of the distribution of correlations \cite{Tranpra92_2015,KnipsnpjQI6_2020}, geometrical threshold criterion \cite{Laskowskipra88_2013}, or interference between multiple copies of the investigated state \cite{Bartkiewiczpra95_2017}. However, all of them were so far limited to mere witnesses of entanglement rather than measures. Entanglement quantification is of considerable interest for both theoretical and practical reasons. Our goal is to combine entanglement measures with the RFI approach and ultimately achieve a reliable RFI entanglement quantification protocol. 
More specifically, we investigate the RFI measure of Bell nonlocality and its relation with entanglement. Although Bell nonlocality and entanglement are distinct resources, one still can establish a direct link between them for specific families of states. Because of this, we restrict our attention to two- and three-qubit states which are of practical importance in quantum information processes. One such example is the family of Werner states which have been instrumental for various important advancements in quantum information \cite{Bennettpra54_1996,Horodeckipra59_1999,Terhalprl85_2000}. Although this family contains examples of states with nonclassical correlations, which nevertheless admit a hidden-variable model, the violation of local-realistic description is still observed for highly entangled cases which are in fact applied in quantum information procedures. We also discuss to what extent the results obtained for the Werner states can be used to estimate the entanglement of other two- and three-qubit states. In other words, we test how precisely one can estimate the entanglement of an unknown state if our RFI approach is applied. Finally, we present an experimental verification of our predictions.


\section{Preliminaries}
\subsection{Entanglement measure}

We now introduce concepts that are relevant to the current investigation.
Let us first consider two-qubit pure state $|\psi\rangle_2$, composed of subsystems $A$ and $B$. The degree of entanglement between both subsystems is given by so-called \textit{concurrence} \cite{Woottersprl80_1998}, $\mathcal{C}(|\psi\rangle_2)=\sqrt{2 \left(1-\textrm{Tr}(\rho^2_A)\right)}$, where $\rho_A$ denotes the reduced density matrix of subsystem $A$. For
mixed states $\rho$ the concurrence is defined by the convex-roof extension \cite{Uhlmannosid5_1998}, 
$\mathcal{C}(\rho)= \min\limits_{\textrm{all decomp.}} \sum_j p_j \mathcal{C}(|\psi_j\rangle)$,
where the minimum average concurrence is taken over all possible convex decompositions $\rho = \sum_j  p_j |\psi_j\rangle \langle \psi_j|$ into pure states. In a special case, when $\rho_2$ denotes two-qubit mixes state, the mixed-state concurrence is given by 
\begin{equation}
\mathcal{C}(\rho_2)= \max\{0,\sqrt{\lambda_1}-\sum_{j=2}^4 \sqrt{\lambda_j}\}
\label{eq:Mixed_concurrence}
\end{equation}
with $\{\lambda_j\}$ being the decreasingly ordered eigenvalues of $\rho_2 (\sigma^y \otimes \sigma^y)\rho_2^T (\sigma^y \otimes \sigma^y)$, where $\sigma^y$ denotes the Pauli matrix and the transposition is performed in any product basis.

The above described measure can be further extended to describe the genuine multipartite entanglement (GME) \cite{Popepra67_2003, LoveQIP6_2007, Mapra83_2011, Chenpra85_2012}, i.e. a scenario when a multipartite state has a minimum amount of entanglement in each bipartition. For instance, if the analyzed pure state $|\psi\rangle_3$ is composed of three subsystems $A$, $B$, and $C$ one can distinguish three bipartitions $\{\gamma|\gamma'\}$, namely $\{A|BC\}$, $\{B|AC\}$, and $\{C|AB\}$. Then, the GME-concurrence is given by \cite{Mapra83_2011}
\begin{equation}
\mathcal{C}_{\textrm{GME}}(|\psi\rangle_3)= \min\limits_{\textrm{all bipart.}} \sqrt{2 \left(1-\textrm{Tr}(\rho^2_{\gamma})\right)},
\label{eq:GME_concurrence}
\end{equation}
where the minimum is taken over all possible bipartitions $\{\gamma|\gamma'\}$ and $\rho_{\gamma}$ denotes the corresponding reduced density matrix of subsystem $\gamma$.
The extension of GME-concurrence to mixed states also follows the convex-roof extension presented above \cite{Mapra83_2011}.

We stress that a general expression for mixed state GME-concurrence still remains unknown. However, it has been successfully evaluated for the so-called X-matrix states \cite{Yuqic7_2007}. These states are represented by a density matrix written in an orthonormal product basis, whose non-zero elements are only the diagonal (denoted by $a_j$ and $b_j$, where $j=\{1,...,2^{N-1}\}$) and/or anti-diagonal elements (given by $z_j$ and its conjugation). The X-matrix states are positive if $|z_j|\leq\sqrt{a_j b_j}$ and we also expect $\sum_j (a_j + b_j) = 1$ to ensure the normalization of $\rho_X$. The GME-concurrence for these states is given by \cite{Hashemipra86_2012}
\begin{equation}
\mathcal{C}_{\textrm{GME}}(\rho_\textrm{X})=2 \max\limits_{i} \{0,|z_i|-\chi_i\},
\label{eq:GME_Xstate}
\end{equation}
where $\chi_i = \sum\limits_{j \neq i} \sqrt{a_jb_j}$.

\subsection{Bell-Nonlocal correlations}
\label{theory_nonlocal}

Next, let us consider an $N$-partite Bell experiment where each party has a choice over two measurement settings $S_i = \{0,1\}$ and each measurement results in one of two possible outcomes $r_i = \{0,1\}$. The corresponding Bell experiment is then fully characterized by the set of joint conditional probability distributions $\textbf{P} = \{P(\vec{r}_N|\vec{S}_N)\}$, where $\vec{r}_N=(r_1,\dots,r_N)$ and $\vec{S}_N=(S_1,\dots,S_N)$. When the participants share a quantum state $\rho$ and the correlations are generated by local measurements performed on their respective subsystems, then $\textbf{P}$ takes the form of $P(\vec{r}_N|\vec{S}_N) = \textrm{Tr} \left(\hat{M}_{r_i|S_i} \bigotimes\limits_{i=1}^N  \rho \right)$, where $\hat{M}_{r_i|S_i}$ is the positive operator-valued measure representing the measurement on the $i$-th party with measurement settings $S_i$.

To make it evident whether a given $\textbf{P}$ can be described by a local realistic description, one can employ a linear function of probabilities called Bell inequality \cite{Bell1964Physique}. It can be written as 
\begin{equation}
\mathcal{I}(\textbf{P})\equiv \sum_{\vec{r}_N,\vec{S}_N} \mu^{\vec{S}_N}_{\vec{r}_N} P(\vec{r}_N|\vec{S}_N) \leq C_{\textrm{LHV}},
\label{eq:BelIn}
\end{equation}
where $\{\mu^{\vec{S}_N}_{\vec{r}_N}\}$ are real coefficients and $C_{\textrm{LHV}}$ refers to the upper threshold of $\mathcal{I}(\textbf{P})$ for the local realistic description. 
Consequently, if one observes a value of $\mathcal{I}(\textbf{P})$ greater than $ C_{\textrm{LHV}}$, the correlations are said to be Bell-nonlocal. 
The value of coefficients $\{\mu^{\vec{S}_N}_{\vec{r}_N}\}$ solely depends on the analyzed model of local realistic description \cite{CHSHprl23_1969,Pitowskypra64_2001,Sliwapla317_2003,Bancalpra88_2013}. For instance, when $N=2$ the above described Bell experiment is characterized by the Clauser-Horne-Shimony-Holt (CHSH) inequality \cite{CHSHprl23_1969}. On the other hand, when $N=3$ the genuine multipartite nonlocal correlations disused in this paper require consideration a set of $185$ Bell inequalities defined in Ref. \cite{Bancalpra88_2013}.

The presence of Bell-nonlocal correlations clearly certifies the presence of entanglement, and this conclusion follows regardless of how $\textbf{P}$ is generated from the underlying state and measurements. Therefore, Eq. \eqref{eq:BelIn} is said to be a device-independent witness for entanglement \cite{Scaraniaps62_2012}. 
To date, the relation between entanglement and Bell-nonlocality has been intensively studied. For instance in Ref.~\cite{Verstraeteprl89_2002} there is shown that $\mathcal{C}(|\psi\rangle_2) = \sqrt{\beta^2_2-1}$, where $\beta_{2}$ stands for the maximal violation of the Clauser-Horne-Shimony-Holt (CHSH) inequality \cite{CHSHprl23_1969}. Similar investigations have been performed for three-qubit states (see for instance \cite{Ghoseprl102_2009,Barasinskisr8_12305_2018} and \cite{Lupra84_2011a} for an experimental demonstration). 

Nevertheless, the above-described demonstration of nonlocal correlations employs carefully chosen measurements whose implementation requires the spatially separated observers to share a complete reference frame and well calibrated devices. Although this assumption is typically made implicit in theoretical works, establishing a common reference frame, as well as aligning and calibrating measurement devices in experimental situations are never trivial tasks. 
Recently, Liang et al. \cite{Liangprl104_2010} have proposed a reference-frame-independent protocol to circumvent the above mentioned problem.
In their approach, the following quantity is considered \cite{Liangprl104_2010,Wallmanpra83_2011}
\begin{equation}
p_{\textrm{V}}(\rho) = \int \omega(\rho, \Omega) d \Omega,
\label{eq:pV}
\end{equation}
where the integration comprises a space of measurement parameters $\Omega$ according to the Haar measure. The function $\omega(\rho, \Omega)$ is an indicator function that takes the value $1$ whenever the generated behavior is nonlocal and $0$ otherwise. What is important, in this approach the nonlocal correlations are quantified without any prior assumptions about specific Bell inequalities \cite{Lipinskanpj_2018,Barasinskipra101_2020,BarasinskiQuantum_2021}. In other words, the generated behavior is nonlocal if at least one inequality of the suitable set of Bell inequalities is violated.
The quantity $p_{\textrm{V}}$, if properly normalized, can be interpreted as a probability of violation of local realism for the measurement operators $\hat{M}_{r_i|S_i}$ sampled randomly according to the Haar measure. To avoid confusion, to describe the quantity $p_{\textrm{V}}$ we prefer to use the unique term nonlocal fraction~\cite{Lipinskanpj_2018}.

\section{Device-independent estimation of entanglement}

\begin{figure}
\centering
\hfill\includegraphics[scale=1.0]{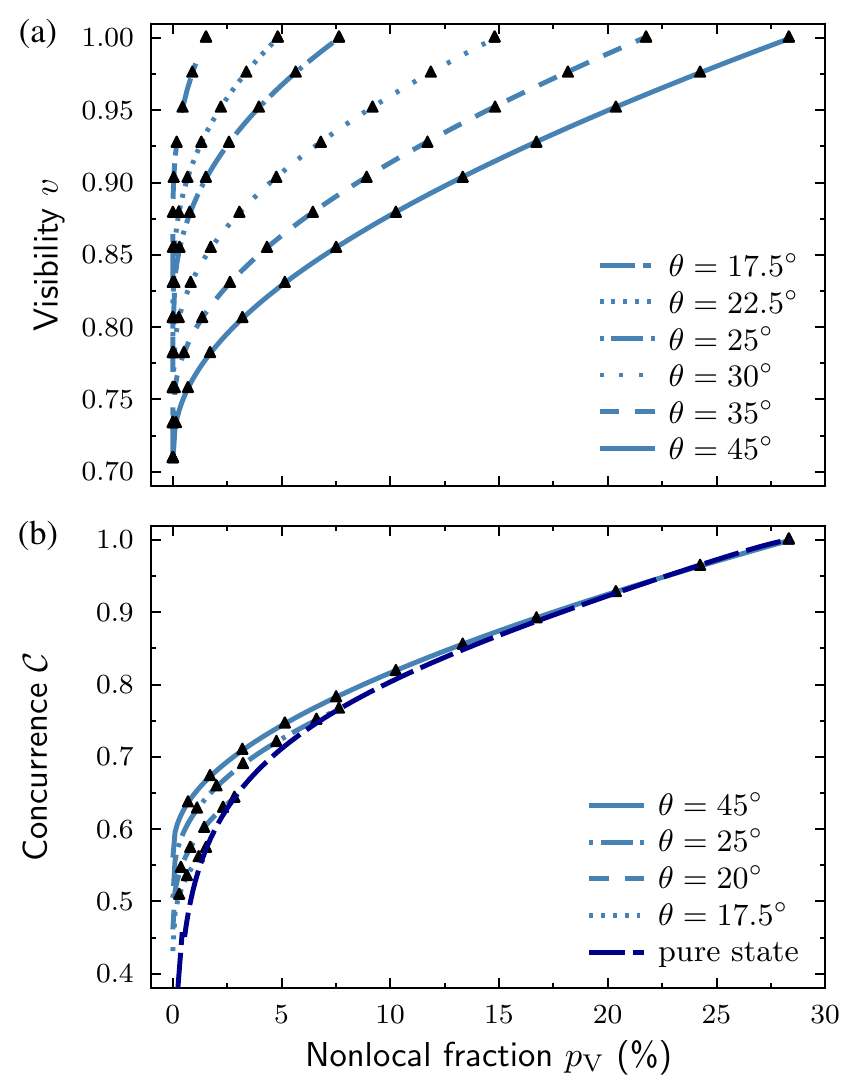}\hspace*{\fill}
\caption{(a) Visibility and nonlocal fraction for  two-qubit Werner-like states given in Eq. \eqref{eq:rho2_theta}. Symbols denote numerical results and solid curves correspond to their analytical approximation in Eq. \eqref{eq:v_Werner2}.
(b) Relation between concurrence $\mathcal{C}$ and nonlocal fraction $p_{\textrm{V}}$ for two-qubit Werner-like states. As previously, symbols denote numerical results while solid curves correspond to analytical approximation.
}
\label{fig:Pure1}
\end{figure}

In this paper we consider a source producing copies of an unknown $N$-qubit state $\rho$, which is transmitted through randomly unitary evolving quantum channels to $N$ local observers. During the $j-$th transmission the state $\rho$ is transformed by $N$ random local unitary operators $U_i^{(j)}$ according to
\begin{eqnarray}
\rho_{\textrm{out}} = U_i^{(j)} \bigotimes\limits_{i=1}^N \rho_{\textrm{in}} \bigotimes\limits_{i=1}^N U_i^{(j)\dag}.
\end{eqnarray}
We assume that the unitary transformation has a timescale that is sufficiently slow to obtain stable measurements for given projections together with their orthogonal counterparts, but the transformation is much faster to apply standard techniques of state analysis~\cite{Jamespra64_2001}. In other words, we can reliably accumulate signal for one particular measurement setting and its orthogonal-projection counterpart but not for all the measurement settings in a row.

We discuss the entanglement assessment protocol of the input state $\rho_{\textrm{in}}$ based on the nonlocal correlations revealed by the output state $\rho_{\textrm{out}}$. As the unitary operators during the $j-$th transmission remain unknown for the observers, the maximal violation of Bell inequalities cannot be determined. Instead, we estimate the nonlocal fraction which is invariant under local unitary transformations applied by each party on the state if one uses the Haar measure for the integration \cite{Lipinskanpj_2018}. 
However, the use of nonlocal fraction has an important disadvantage which is the lack of analytical solutions \cite{Liangprl104_2010,Lipinskanpj_2018}, and so, the numerical calculations are used to determine the nonlocal fraction. 

\subsection{Quantifying Bipartite Entanglement}

\subsubsection{Two-qubit Werner-like states}
First, we consider the scenario when the input state is given in a form of an arbitrary two-qubit pure state $\ket{\theta}_2=\cos{\theta}\ket{00} + \sin{\theta}\ket{11}$ subjected to white noise:
\begin{eqnarray}
\rho_2(\theta,v) = v \ket{\theta}_2 \bra{\theta}+\frac{1-v}{4} \id_4,
\label{eq:rho2_theta}
\end{eqnarray}
where $\id_4$ is the $4\times4$ identity matrix, $v$ stands for the state visibility ($0<v\leq1$), and we assume without loss of generality that $0<\theta\leq 45^{\circ}$. The concurrence is given by
\begin{eqnarray}
\mathcal{C}(\rho_2) &=& \frac{ v\Big( 2 \sin(2 \theta) + 1 \Big) - 1}{2}.
\label{eq:conc_Werner2}
\end{eqnarray}

Such states play an important role in quantum information theory as they directly refer to the states generated at the output of the nonlinear process designed in real experiments based on entangled photons \cite{Kwiatprl75_1995,TorresPO56_2011}. In this context, the white noise which enters Eq. \eqref{eq:rho2_theta} is a first approximation of the imperfections~~occurring in the experimental setup.

A particular example of the states in Eq.~\eqref{eq:rho2_theta} is the two-qubit Werner state \cite{Wernerpra40_1989}, $\rho^\textrm{W}_2(v)=\rho_2(\theta=45^{\circ},v)$ \cite{Bennettpra54_1996,Horodeckipra59_1999,Terhalprl85_2000,Horodecki2009RevModPhys}. For the Werner states concurrence depends only on the visibility, $\mathcal{C}(\rho^\textrm{W}_2) = \frac{3 v - 1}{2}$. Therefore, the estimation of this parameter is equivalent to the entanglement measurement. 

To do that we calculate the nonlocal fraction. Note that the nonlocal correlations of two-qubit states are fully characterized by the CHSH inequality, assuming the freedom in relabeling all measurement settings and/or outcomes and/or parties \cite{MasanesQIC3_2003,Collins_2004}. By straightforward calculations (see Appendix \ref{appendixA}) one can show that $p_{\textrm{V}}$ of the Werner state is
\begin{eqnarray}
p_{\textrm{V}}(v)=\frac{2 \Big((1-v^2) \arctan(\frac{\sqrt{2 v^2-1}}{1-v^2})-3 \sqrt{2 v^2-1}\Big)}{v^2},\nonumber\\
\label{eq:pV_Werner2}
\end{eqnarray}
which is a monotonic function of $v$. In other words, a direct measurement of $p_{\textrm{V}}$ allows to estimate visibility, and hence, the value of the concurrence $\mathcal{C}(\rho^\textrm{W}_2)$.

Naturally, for general state \eqref{eq:rho2_theta} the nonlocal fraction depends on both the visibility $v$ and angle $\theta$ (see Fig. \ref{fig:Pure1}(a)). Although the analytical solution of $p_{\textrm{V}}$ remains unknown in this case, one can always find its approximation. In particular, one can establish the visibility $v$ by 
\begin{eqnarray}
v(\theta, p_{\textrm{V}})=v_2^{\textrm{cr}}(\theta)+f_1(\theta)~p_{\textrm{V}}^{1/4} + f_2(\theta)~p_{\textrm{V}}^{1/2} +f_3(\theta)~p_{\textrm{V}},\nonumber\\
\label{eq:v_Werner2}
\end{eqnarray}
where 
\begin{eqnarray}
f_1(\theta) &=& (0.19674 - 1.3982~\theta + 4.712274~\theta^2 \nonumber\\ &-& 6.7193~\theta^3+ 3.3384~\theta^4)/\sqrt{10},\nonumber\\
f_2(\theta) &=& 0.11886 - 0.011544~\theta^{-1} - 0.363104~\theta \nonumber\\ &+& 0.460436~\theta^2- 0.204953~\theta^3,\nonumber\\
f_3(\theta) &=& (0.03848 - 0.011~\theta^{-1} - 0.02531~\theta \nonumber\\ &-& 0.018331~\theta^2+ 0.017373~\theta^3)\cdot10^{-2},\nonumber
\end{eqnarray}
and $v_2^{\textrm{cr}}(\theta)=1/\beta_2$ denotes the critical visibility with the maximal violation of the CHSH inequality $\beta_2= (\sin^2(2 \theta) + 1)^{1/2}$ \cite{Verstraeteprl89_2002}. 

\begin{figure}
\centering
\hfill\includegraphics[scale=1.]{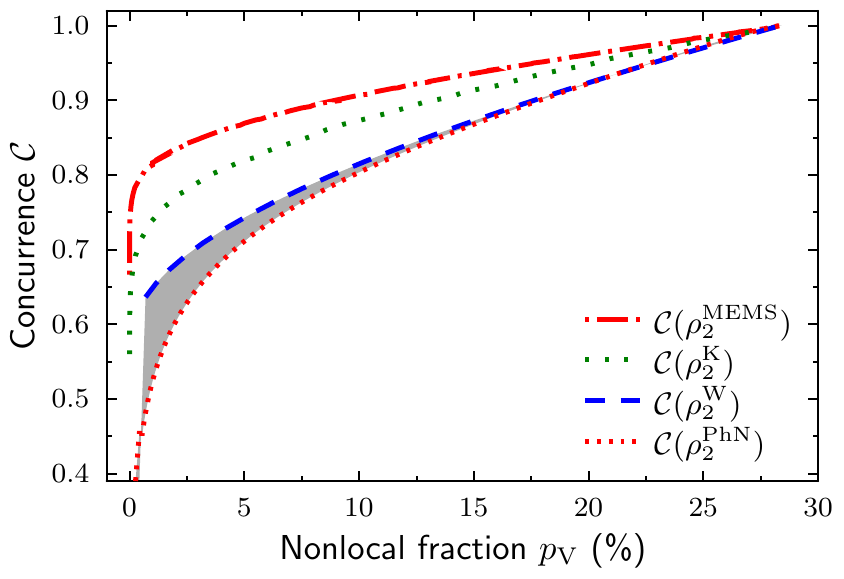}\hspace*{\fill}
\caption{The region of possible values of concurrence for given nonlocal fraction. The grey region corresponds to two-qubit GHZ symmetric mixed states and the four curves represent maximally entangled mixed states (red dashed-dotted curve), Kagalwala states \cite{exampleC} (green dotted curve), Werner states (blue dashed curve), two-qubit GHZ state subjected to the local phase-damping noise (red dotted curve).
}
\label{fig:Mixed2}
\end{figure}

As presented in Fig. \ref{fig:Pure1}(a), this approximation provides a good agreement with our numerical results. Therefore, substituting Eq. \eqref{eq:v_Werner2} into Eq. \eqref{eq:conc_Werner2} one obtains the concurrence $\mathcal{C}(\rho_2)$ depending on the angle $\theta$ and the nonlocal fraction $p_{\textrm{V}}$ (see Fig. \ref{fig:Pure1}(b)). 
Based on these outcomes, the following remarks can be drawn: 

(i) Whenever an observed $p_{\textrm{V}} \geq 7\%$ the difference between $\mathcal{C}(\rho^\textrm{W}_2)$ and $\mathcal{C}(\rho_2)$ (hereinafter $\Delta_2^\textrm{W}$) is not greater than $0.02$ and vanishes when $p_{\textrm{V}}$ increases. It means, that the concurrence $\mathcal{C}(\rho_2)$ can be estimated (with precision $\Delta_2^\textrm{W}$) assuming that $\rho_2 \equiv \rho^\textrm{W}_2$.

(ii) For $p_{\textrm{V}} < 7\%$, the above remark is still valid if $\theta \geq 25^{\circ}$ and $p_{\textrm{V}}\geq0.5\%$. In other words, the angle $\theta$ is meaningless in such regime and the concurrence can be estimated on $\mathcal{C}(\rho^\textrm{W}_2)$. For other cases, the difference $\Delta_2^\textrm{W}$ increases for decreasing angle $\theta$.

(iii) Finally, Eq. \eqref{eq:v_Werner2} can be used to established the lower bound of $\mathcal{C}(\rho_2)$ vs. $p_{\textrm{V}}$. Specifically, for a given value of the nonlocal fraction there exists such angle $\theta_0$ so that the visibility $v(\theta_0,p_{\textrm{V}})$ in Eq. \eqref{eq:v_Werner2} is equal to $1$. Then, the lower bound is given by $\mathcal{C}(\rho_2) \geq \sin(2 \theta_0)$ and the equality is provided by the pure state $\ket{\theta_0}_2$. The lower bound can be approximated by
\begin{eqnarray}
\mathcal{C}(\ket{\theta_0}_2) = \frac{0.6784}{\sqrt{10}}~p_{\textrm{V}}^{1/4} - 1.59\cdot10^{-2}~p_{\textrm{V}}^{1/2} + 10^{-4}~p_{\textrm{V}}.\nonumber\\
\label{eq:c_theta}
\end{eqnarray}
Based on this result, one can find that the difference $\Delta_2^\textrm{W}<0.164$ for an arbitrary  angle $\theta$ and $p_{\textrm{V}}\geq0.5\%$.


\subsubsection{General two-qubit mixed states}

In order to present the usefulness of our entanglement-assessment protocol for a broader range of two-qubit state $\rho_{\textrm{in}}$, we will now consider two examples where we apply our protocol.

\textit{Example 1: Two-qubit GHZ symmetric mixed state (GSMS)~--} These states represent the entire family of two-qubit mixed states with the same symmetry as the two-qubit GHZ state $\ket{45^{\circ}}_2$ \cite{Eltschkaprl108_2012}. For instance, the Werner states $\rho^\textrm{W}_2$ but also the $\ket{45^{\circ}}_2$ state subjected to the local phase-damping or depolarizing noise \cite{Krauslnp_1983}. 
The GHZ symmetric states are defined as \cite{Eltschkaprl108_2012} 
\begin{eqnarray}
\rho^{\textrm{GSMS}}_2(x,y) &=& (\sqrt{2} y+x) \ket{45^{\circ}}_2\bra{45^{\circ}} \nonumber\\
&+&(\sqrt{2} y-x) \ket{-45^{\circ}}_2\bra{-45^{\circ}}+\frac{1-2\sqrt{2} y}{4} \id_2,\nonumber
\end{eqnarray}
where $|y|\leq (2 \sqrt{2})^{-1}$ and $|x| \leq (1 + 2 \sqrt{2} y)/4$. Using Eq. \eqref{eq:Mixed_concurrence} one gets the concurrence $\mathcal{C}(\rho^{\textrm{GSMS}}_2)= \max\{0,2 |x|+\sqrt{2}y-1/2\}$. 

Next, the relation between $\mathcal{C}(x,y)$ and the nonlocal fraction for $10^4$ randomly generated GSMS states has been analyzed. As a result (Fig. \ref{fig:Mixed2}), we have found that the upper bound of such relation is provided by the Werner states $\rho_2^\textrm{W} \equiv \rho^{\textrm{GSMS}}_2(v/2,v/(2\sqrt{2}))$. The lower bound, on the other hand, is established by the maximally nonlocal mixed states, i.e. Bell diagonal states which produce a maximal value of $\beta_2$ for given concurrence \cite{Batlejpa44_2011}. These states are given by
\begin{eqnarray}
\rho^{\textrm{PhN}}_2(x) &=& \frac{1+2x}{2} \ket{45^{\circ}}_2\bra{45^{\circ}} + \frac{1-2x}{2} \ket{-45^{\circ}}_2\bra{-45^{\circ}},\nonumber
\end{eqnarray}
and 
describe the $\ket{45^{\circ}}_2$ state subjected to the local phase-damping noise \cite{Krauslnp_1983}. The relation between the concurrence and the nonlocal fraction in this case is given by $\mathcal{C}(\rho^{\textrm{PhN}}_2) = \mathcal{C}(\ket{\theta_0}_2)$ written in Eq. \eqref{eq:c_theta}. Therefore, if one knows the nonlocal fraction of an arbitrary GHZ symmetric state then its concurrence is limited by $\mathcal{C}(\ket{\theta_0}_2)\leq \mathcal{C}(\rho^{\textrm{GSMS}}) \leq \mathcal{C}(\rho^{\textrm{W}}_2)$. This limitation is of great importance if the remarks drown in the previous subsection are taken into account. This is, the concurrence of an arbitrary GHZ symmetric state can be determined with accuracy not greater than $\Delta_2^\textrm{W}$ if the measured $p_{\textrm{V}} \geq 7\%$.
Note that, in general, the GSMS may denotes the experimentally generated state $\ket{45^{\circ}}_2$  subjected to an unknown source of noise if such noise does not change the symmetry of the input state.

\begin{figure}
\centering
\hfill\includegraphics[scale=1.0]{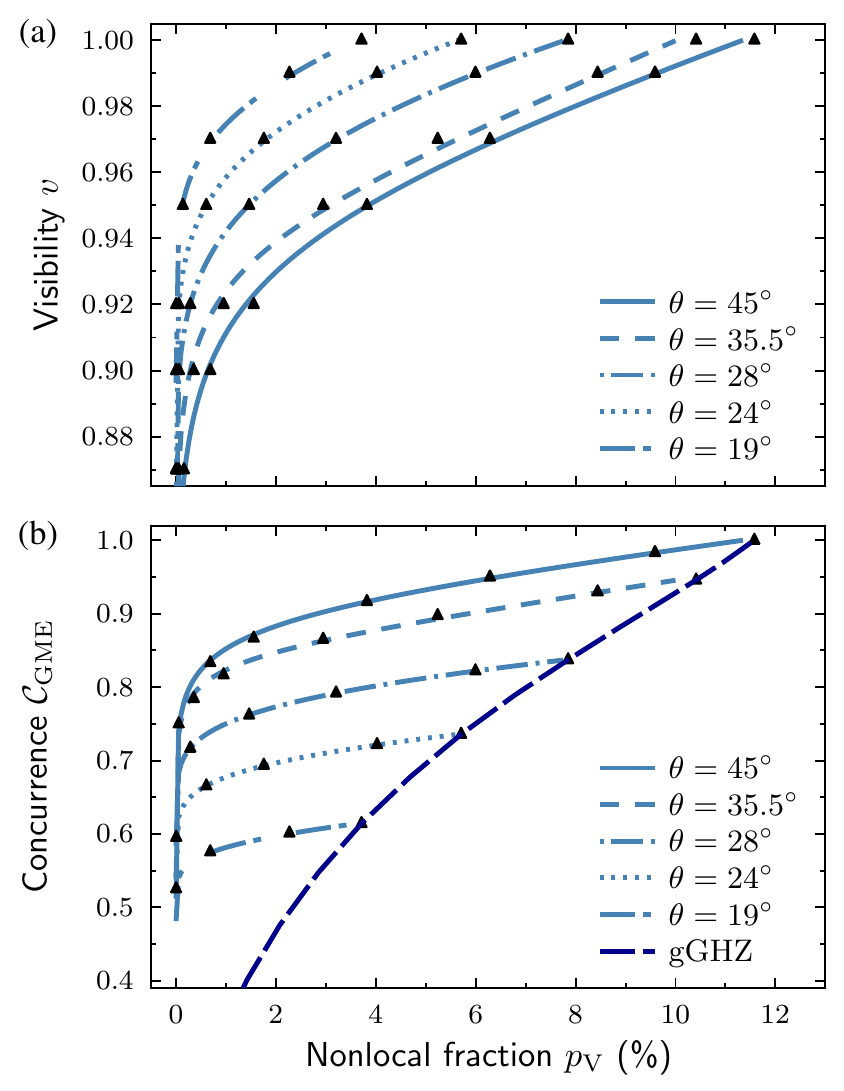}\hspace*{\fill}
\caption{(a) Visibility and nonlocal fraction for three-qubit Werner-like states given in Eq. \eqref{eq:rho3_theta}. Symbols denote numerical results and solid curves correspond to their analytical approximation in Eq. \eqref{eq:v_Werner3}.
(b) Relation between genuine concurrence $\mathcal{C}_{\textrm{GME}}$ and nonlocal fraction $p_{\textrm{V}}$ for three-qubit Werner-like states. As previously, symbols denote numerical results while solid curves correspond to analytical approximation.
}
\label{fig:Pure3}
\end{figure}

\textit{Example 2: Maximally entangled mixed state (MEMS)~--} As a final example we consider the states which maximize the value of the concurrence for a given value of the violation of the CHSH inequality \cite{Munropra64_2001,Weipra67_2003}
\begin{eqnarray}
\rho^{\textrm{MEMS}}_2(\gamma) &=& \gamma \ket{45^{\circ}}_2\bra{45^{\circ}} + (1-\gamma) \ket{01}\bra{01},\nonumber
\end{eqnarray}
where $\frac{2}{3}\leq \gamma \leq 1$. Based on numerical calculation we have found that
\begin{eqnarray}
\mathcal{C}(\rho^{\textrm{MEMS}}_2) &=& 1/\sqrt{2} + 0.1125/\sqrt{10}~p_{\textrm{V}}^{1/4} -  9.0\cdot10^{-4}~p_{\textrm{V}}^{1/2}\nonumber\\
&+& 2.83\cdot10^{-5}~p_{\textrm{V}}.\nonumber
\end{eqnarray}
As we see in Fig. \ref{fig:Mixed2}, the concurrence $\mathcal{C}(\rho^{\textrm{MEMS}}_2)$ exceeds $\mathcal{C}(\rho^{\textrm{W}}_2)$ in the entire range of $p_{\textrm{V}}$. However, the difference between these two quantities is not greater that $0.173$. 

Finally, our numerical calculations performed for randomly generated two-qubit mixed states $\rho$ always satisfied the relation 
\begin{eqnarray}
\mathcal{C}(\ket{\theta_0}_2)\leq\mathcal{C}(\rho) \leq \mathcal{C}(\rho^{\textrm{MEMS}}_2),
\end{eqnarray}
if they reveal the same value of $p_{\textrm{V}}$. Therefore, we conjecture that the MEMS and pure states $\ket{\theta}_2$ provide an upper and lower limit for $\mathcal{C}(\rho)$ vs $p_{\textrm{V}}$ for two-qubit states. 

\subsection{Quantifying Genuine Tripartite Entanglement}

\subsubsection{Three-qubit Werner-like states}

Now we proceed to estimate the genuine multipartite entanglement. We follow the same procedure as before, i.e. we analyze the relationship between the GME-concurrence and nonlocal fraction. 
First, we concentrate on the three-qubit Werner-like states which serve as a benchmark for the robustness of multipartite entanglement \cite{Durpra61_2000}
\begin{eqnarray}
\rho_3(\theta,v) = v \ket{\theta}_3\bra{\theta}+\frac{1-v}{8} \id_8,
\label{eq:rho3_theta}
\end{eqnarray}
where $\ket{\theta}_3=\cos{\theta}\ket{000} + \sin{\theta}\ket{111}$ is the generalized GHZ state (gGHZ) and $\id_8$ is the $8 \times 8$ identity matrix denoting the presence of the white noise. As before, $v$ stands for the state visibility ($0<v\leq1$) and we assume  $0<\theta\leq 45^{\circ}$. Using Eq. \eqref{eq:GME_Xstate} one can find the GME-concurrence as
\begin{eqnarray}
\mathcal{C}_{\textrm{GME}}(\rho_3) &=& \frac{\Big( 3 \sin(2 \theta) + 2 \Big) v - 2}{3}.
\label{eq:concW3}
\end{eqnarray}

\begin{figure}
\centering
\hfill\includegraphics[scale=1.0]{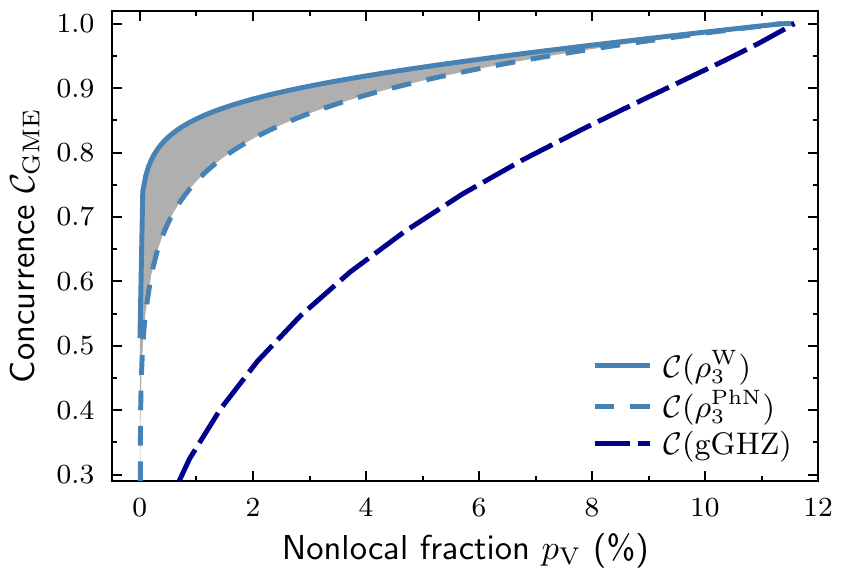}\hspace*{\fill}\\
\caption{The region of possible values of genuine concurrence for given nonlocal fraction. The grey region corresponds to three-qubit GHZ symmetric mixed states and the three curves represent Werner states (solid curve), three-qubit GHZ state subjected to the local phase-damping noise (dotted curve), and generalized GHZ states (dashed curve).}
\label{fig:Mixed3}
\end{figure}

In order to certify the genuine multipartite entanglement, we estimate the nonlocal fraction for the genuine multipartite nonlocal correlations. Such an estimation requires testing all $185$ families of  Bell inequalities (see \cite{Barasinskipra101_2020}). As a result (Fig \ref{fig:Pure3}(a)), we have found that the visibility $v$ in Eq. \eqref{eq:rho3_theta} can be approximated by $p_{\textrm{V}}$ using
\begin{eqnarray}
v(\theta,p_{\textrm{V}})=v_3^{\textrm{cr}}(\theta)+g_1(\theta)~p_{\textrm{V}}^{1/6} + g_2(\theta)~p_{\textrm{V}}^{1/2} +g_3(\theta) ~p_{\textrm{V}},\nonumber\\
\label{eq:v_Werner3}
\end{eqnarray}
where the critical visibility $v_3^{\textrm{cr}}(\theta) = 1/\beta_3$ and 
\begin{eqnarray}
\beta_3 =\left\{ \begin{array}{ll}
1 + 0.0622 \theta + 1.697 \theta^2  & \textrm{for $0\leq \theta < 14.94^{\circ}$}\\
− 3.391 \theta^3 + 1.442 \theta^4 & \\
\left(1 + 2 \sqrt{1 + \sin^2(2 \theta)}\right)/3 & \textrm{for $14.94^{\circ}\leq \theta < 29.5^{\circ}$}\\
\sqrt{2 \sin^2(2 \theta)} & \textrm{for $29.5^{\circ}\leq \theta < 45^{\circ}$}
\end{array} \right.\nonumber
\end{eqnarray}
is the maximal strength of Bell-nonlocality for three-qubit Werner-like states (see \cite{Barasinskiprl122_2019}). The other functions which ente Eq. \eqref{eq:v_Werner3} are given by 
\begin{eqnarray}
g_1(\theta) &=& \max\{-0.061297 + 0.55512~\theta - 0.42815~\theta^2,\nonumber\\&& -18.58393 + 57.9917~\sqrt{\theta} - 50.2727~\theta \nonumber\\&&+ 11.209~\theta^2\}/10^{1/3},\nonumber\\
g_2(\theta) &=& \min\{0, 0.76306 - 4.13852~\theta + 8.28077~\theta^2\nonumber\\&& - 7.2943~\theta^3 + 2.38884~\theta^4\},\nonumber\\
g_3(\theta) &=& \max\{0.0001151 - 0.0004063~\theta + 0.0004321~\theta^2,\nonumber\\&& -0.015237 + 0.084803~\theta - 0.17408~\theta^2\nonumber\\&& + 0.15723~\theta^3 - 0.052804~\theta^4\}.\nonumber
\end{eqnarray}

Based on Eq. \eqref{eq:concW3} and \eqref{eq:v_Werner3}, the GME-concurrence has been obtained as a function of $p_{\textrm{V}}$. As we see in Fig. \ref{fig:Pure3}(b), in contrast to $\rho_2(\theta,v)$, here the angle $\theta$ is meaningful in the entire range of attainable $p_{\textrm{V}}$. For instance, if one takes $\theta_1=45^{\circ}$ (i.e. the three-qubit Werner state) and $\theta_2=35^{\circ}$, the GME-concurrence is explicitly written as
\begin{eqnarray}
\mathcal{C}_{\textrm{GME}}(\theta_1) &=& 0.512 + 0.186~p_{\textrm{V}}^{1/6} - 7.1\cdot10^{-3}~p_{\textrm{V}}^{1/2} \nonumber\\&&+ 1.12\cdot10^{-4}~p_{\textrm{V}},\nonumber\\
\mathcal{C}_{\textrm{GME}}(\theta_2) &=& 0.542 + 0.155~p_{\textrm{V}}^{1/6} - 8.2\cdot10^{-3}~p_{\textrm{V}}^{1/2} \nonumber\\&&+ 1.52\cdot10^{-4}~p_{\textrm{V}}.
\label{eq:C_Werner3b}
\end{eqnarray}
Using these equations one can easily find the difference $\Delta^{\textrm{W}}_3 = \mathcal{C}_{\textrm{GME}}(\theta_1)-\mathcal{C}_{\textrm{GME}}(\theta_2) \geq 0.04$ when $p_{\textrm{V}}>2\%$ and it grows for higher $p_{\textrm{V}}$. 
Therefore, in order to establishes GME-concurrence $\mathcal{C}_{\textrm{GME}}(\rho_3)$, it is needed to evaluate not only the value of $p_{\textrm{V}}$ but also the underlying angle $\theta$.
Without prior knowledge of the angle $\theta$, its value can be determined from the distribution of the strength of violation for random measurements (Appendix \ref{appendixB}). The latter one requires the accumulation of data on the strength of violation of local realism for sequence of randomly chosen measurements. In a typical experimental investigation of $p_{\textrm{V}}$ \cite{Shadbolt2012SciRep,Barasinskipra101_2020,BarasinskiQuantum_2021} such a set is known without any additional effort. 

On the other hand, by inserting $v(\theta,p_{\textrm{V}})=1$ in Eq. \eqref{eq:v_Werner3} one can derive the GME-concurrence for pure states $|\theta\rangle_3$. It can be approximated by \cite{Barasinskipra101_2020} 
\begin{eqnarray}
\mathcal{C}_{\textrm{GME}}(\ket{\theta}_3) &=& \left(0.068~p_{\textrm{V}} + 0.06~p_{\textrm{V}}^{1/2}\right)^{1/2},
\label{eq:C_GHZ3}
\end{eqnarray}
which denotes the lower bound of $\mathcal{C}_{\textrm{GME}}(\rho_3)$ with given $p_{\textrm{V}}$.

\subsubsection{Other mixed states}

\textit{Example 3: Three-qubit GHZ symmetric mixed state (GSMS)~--}
A natural extension of the three-qubit Werner-like states is the family of GHZ symmetric states. In the  three-qubit case, they are given by
\begin{eqnarray}
\rho^{\textrm{GSMS}}_3(x,y) &=& \left(\frac{2 \sqrt{3}}{3} y+x\right) \ket{45^{\circ}}_3\bra{45^{\circ}} \\
+\Bigg(\frac{2 \sqrt{3}}{3} y &-& x \Bigg) \ket{-45^{\circ}}_3\bra{-45^{\circ}}+\frac{3-4\sqrt{3} y}{24} \id_8,\nonumber
\end{eqnarray}
where $\frac{-1}{4\sqrt{3}} \leq y \leq \frac{\sqrt{3}}{4}$, $|x| \leq (1 + 4 \sqrt{3} y)/8$ and the GME-concurrence $\mathcal{C}_{\textrm{GME}}(\rho^{\textrm{GSMS}}_3) = \max\{0,2 |x|+\sqrt{3}y-3/4\}$

For these states similar remarks can be drawn as in \textit{Example 1}. Specifically, the upper bound of the GME-concurrence for given value of $p_{\textrm{V}}$ is provided by the three-qubit Werner state $\rho^{\textrm{W}}_3$. The lower bound is observed for $\rho^{\textrm{PhN}}_3(x) = \left(\frac{1}{2}+x\right) \ket{45^{\circ}}_3\bra{45^{\circ}} +\left(\frac{1}{2}-x\right) \ket{-45^{\circ}}_3\bra{-45^{\circ}}$, i.e. the GHZ state subjected to the local phase-damping noise \cite{Krauslnp_1983}. Interestingly, results obtained for $\rho^{\textrm{PhN}}_3$ are significantly different with respect to those of $\ket{\theta}_3$, as opposed to the case of two-qubit. The GME-concurrence is approximated by:
\begin{eqnarray}
\mathcal{C}_{\textrm{GME}}(\rho^{\textrm{PhN}}_3) = 0.4012~p_{\textrm{V}}^{1/6} - 0.0118~p_{\textrm{V}}^{1/2}+ 9.0\cdot10^{-5}~p_{\textrm{V}}, \nonumber
\end{eqnarray}
and hence, the difference $\Delta^W_3 \leq 0.14$. 
In summary, for all $\rho^{\textrm{GSMS}}_3$ states the following relation is observed 
\begin{eqnarray}
\mathcal{C}(\ket{\theta}_3) < \mathcal{C}(\rho^{\textrm{PhN}}_3)\leq \mathcal{C}(\rho^{\textrm{GSMS}}_3) \leq \mathcal{C}(\rho^{\textrm{W}}_3),
\end{eqnarray}
where we assume that each state reveals the same value of the nonlocal fraction.

\begin{figure}
\centering
\hfill\includegraphics[scale=0.8]{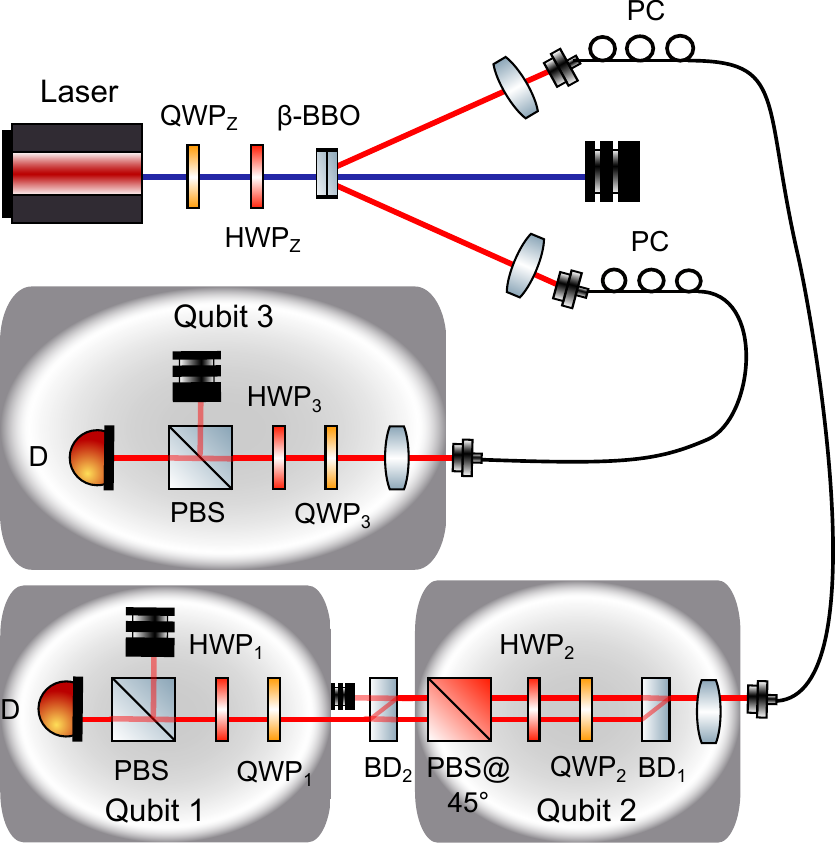}\hspace*{\fill}
\caption{Experimental setup. Legend: PBS -- polarization beam splitter, BD -- beam displacer, PC -- polarization controller, $\beta$-BBO -- non-linear crystal $\beta$-barium borate, D -- detector, HWP -- half-wave plate, QWP -- quarter wave-plate.}
\label{fig:Exsetup}
\end{figure}

\section{Experimental Implementation}
\subsection{Experimental setup}


\begin{figure}
\centering
\hfill\includegraphics[scale=0.8]{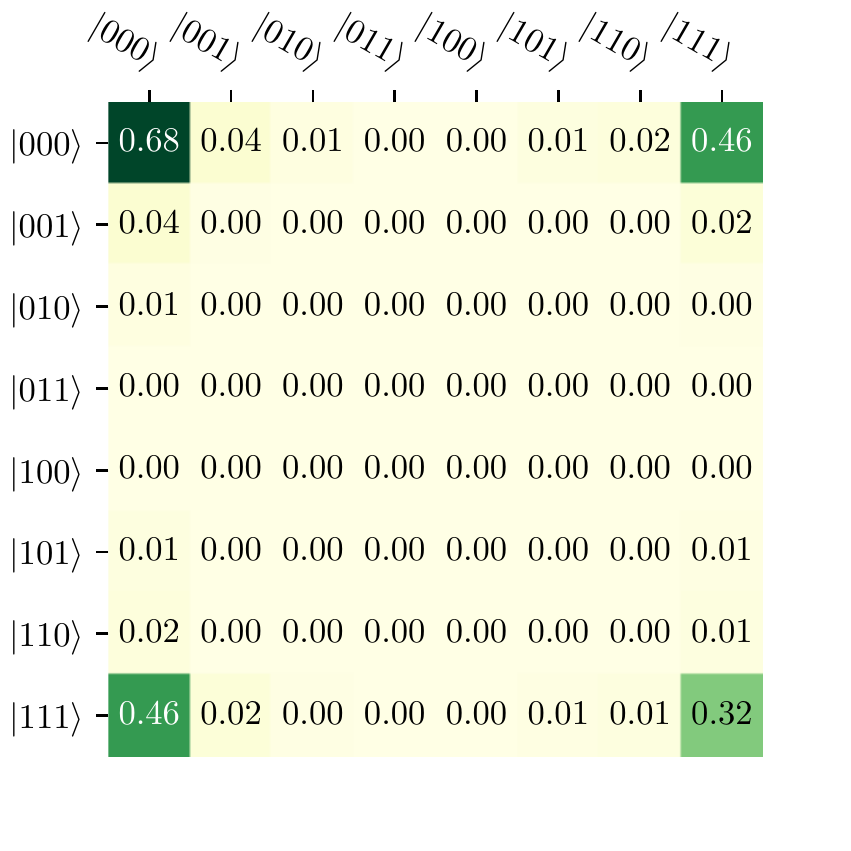}\hspace*{\fill}
\caption{Visualization of real part of the density matrix $\rho^{\textrm{expt}}$ $\theta = 35^\circ$. All values of the imaginary part of $\rho^{\textrm{expt}}$ were less than $0.025$.}
\label{fig:denMat}
\end{figure}

We have constructed the experimental setup depicted in Fig.~\ref{fig:Exsetup} to produce and characterise three-qubit states. Our experiment is implemented on the platform of linear optics and it encodes qubits into spatial and polarisation states of single photons. The setup utilises entangled photon pairs generated using Type I parametric down-conversion in a $\beta$-BBO crystal cascade (referred to as Kwiat source~\cite{Kwiat1999PRA}) at $\lambda = 710$~nm. A laser beam of  a wavelength of $\lambda = 355$~nm pumps two identically cut non-linear crystals, optical axis in mutually perpendicular planes defining horizontal and vertical basis. If pumped by horizontally (vertically) polarised pump beam, pairs of vertically (horizontally) polarised photons are generated. By setting half-wave plate HWP$_{\text{Z}}$ at angle $\tfrac{\theta}{2}$ both crystals are coherently pumped and generate photons in a state of the form of
\begin{equation}
\cos{\theta} \ket{HH} + \sin{\theta} \ket{VV} \,.\label{eq:prep}
\end{equation}
The first (second) position in the ket stands for polarization of the first (second) photon, respectively. Probability of generating two pairs simultaneously is negligible.


In order to generate the three-qubit states we incorporate spatial mode encoding to be used in addition to polarisation encoding. For this purpose the first photon is subjected to the beam displacer (BD$_{1}$). BD$_{1}$ deviates vertically polarised photons upwards whereas horizontally polarised photons continue straightforward. Therefore one can denote by $\ket{0}$ ($\ket{1}$) spatial mode of photons in the upper (lower) arm. At the same time by associating $H$ ($V$) polarisation with logical states $\ket{0}$ ($\ket{1}$) one can immediately identify that by the action of BD$_{1}$ the original two-qubit state~\eqref{eq:prep} becomes a generalised GHZ state in its canonical form 
\begin{equation}
\ket{\theta}_3 = \cos{\theta}\ket{000} + \sin{\theta}\ket{111}\, .\label{eq:gghz}
\end{equation}
Here the first qubit in the ket denotes first photon's spatial mode and second (third) qubit stands for first (second) photon's polarisation state.

Having the desired state prepared, all 3 qubits are subjected to local projections (hereafter $\ket{\hat{\Pi}_1 \otimes \hat{\Pi}_2 \otimes \hat{\Pi}_3}$). The third qubit is projected simply by using a combination of quarter and half wave plates (QWP$_{3}$ and HWP$_{3}$) accompanied by polarising beam splitter (PBS). The remaining two qubits are encoded into spatial and polarisation state of the first photon. Using a similar sequence (QWP$_{2}$, HWP$_{2}$ and PBS) spreading over both spatial modes of this photon we achieve projection of the second qubit. At this stage a BD$_{2}$ is used to convert the spatial encoding of the first qubit to polarisation encoding. Once polarisationally encoded the sequence of QWP$_{1}$,  HWP$_{1}$ and PBS is used to perform first qubit's projection. At the end of the setup, both photons are led to single-photon detectors and the rate of coincident detections is measured for every projection setting.

For the purposes of this experiment, we require the setup to prepare and characterise all pure computational basis states, i.e. $\ket{\textrm{basis}} = \{\ket{000}, \ket{001}, \dots,\ket{111}\}$. This is simply achieved by setting $\theta = 0^{\circ}$ resulting in generation of the $\ket{000}$ state and imposing single-qubit NOT gates in the modes where the qubit is required in the $\ket{1}$ state. These NOT gates are implemented by adding a $45^{\circ}$ bias to the HWP associated with this qubit. All these states were later used to synthesise white noise. After that, various quasi-pure GHZ states were also prepared. All experimental data accumulated in this experiment are available in Supplement~\cite{supplement3}.

\subsection{Nonlocal fraction measurements - aligned reference frames}

\begin{figure}
\centering
\hfill\includegraphics[scale=1.]{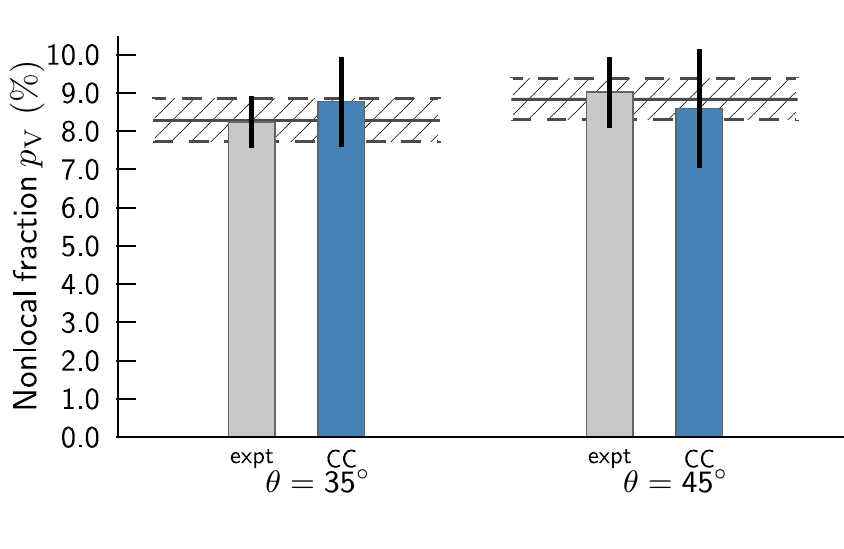}
\hspace*{\fill}
\caption{Comparison of the nonlocal fraction estimated for the reconstructed density matrix $p_{\textrm{V}}(\rho^{\textrm{expt}}_{\theta})$ (gray bar) and measured coincidence detections $p_{\textrm{V}}^{\textrm{CC}}(\theta)$ (blue bar) for angle $\theta = 35^{\circ}, 45^{\circ}$. Horizontal hatched areas stand for theoretical predictions, where solid lines correspond to $p_{\textrm{V}}(\theta,v_{\theta})$ while dashes lines denotes $p_{\textrm{V}}(\theta,v_{\theta}\pm0.03)$.}
\label{fig:comparison_pV}
\end{figure}

First, we consider a scenario when the observers share common reference frames. The experimental setup has been adjusted in such a way to generate the gGHZ states, $\ket{\theta}_3$, for two different angles accounting for $35^{\circ}$ and $45^{\circ}$. Note that the later case denotes the prototype GHZ state. For each adjustment of $\theta$, the output-state density matrix, $\rho^{\textrm{expt}}_{\theta} \equiv \rho^{\textrm{expt}}_3(\theta)$, has been reconstructed by evaluating the quantum state tomography and maximum-likelihood estimation~\cite{Halenkova2012ApplOpt,Hradil2004}. 
An exemplary result is shown in Fig. \ref{fig:denMat}. Then, we determined the fidelity $F$ of $\rho^{\textrm{expt}}_{\theta}$ with respect to the ideal pure state $|\theta\rangle_3$, $F(\rho^{\textrm{expt}}_{\theta})= \textrm{Tr}(\rho^{\textrm{expt}}_{\theta}|\theta\rangle_3\langle \theta|)$. As a result we have found that $F(\rho^{\textrm{expt}}_{\theta})$ is always greater than $0.980 \pm 0.002$ for all values of $\theta$ confirming a good quality of our source. The uncertainty of the fidelity has been determined by Monte Carlo simulations of Poissonian noise distribution.

\begin{figure}
\centering
\hfill\includegraphics[scale=1.]{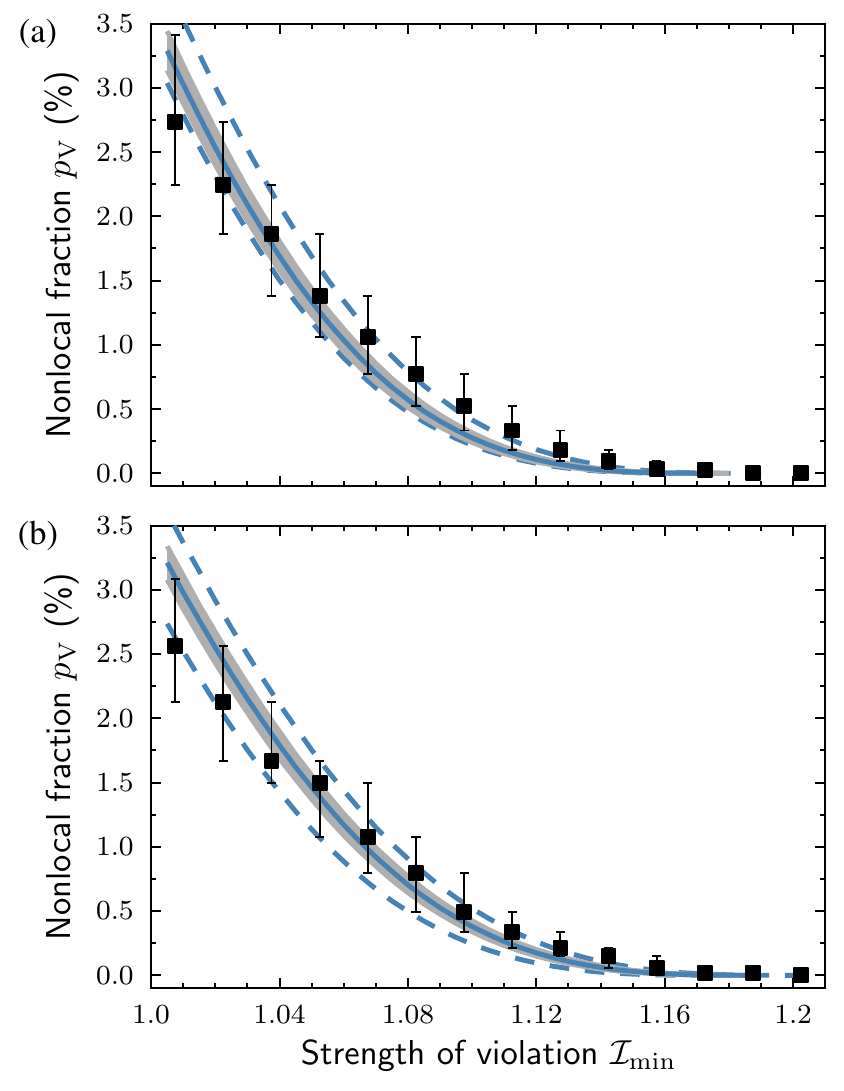}\hspace*{\fill}
\caption{Distribution of the strength of violation for randomly sampled measurements for (a) $\theta = 35^{\circ}$, and (b) $\theta = 45^{\circ}$. In both panels symbols denote experimental results while gray areas depict theoretical predictions for $\rho_3(\theta,v_{\theta}\pm0.003)$. Dashed lines correspond to theoretical calculations for (a) $\rho_3(35^{\circ},1)$ and $\rho_3(35^{\circ},0.985)$ (b) $\rho_3(45^{\circ},0.990)$ and $\rho_3(45^{\circ},0.975)$.}
\label{fig:Nonlocal_dist}
\end{figure}

\begin{figure*}
\centering
\hfill\includegraphics[scale=0.9]{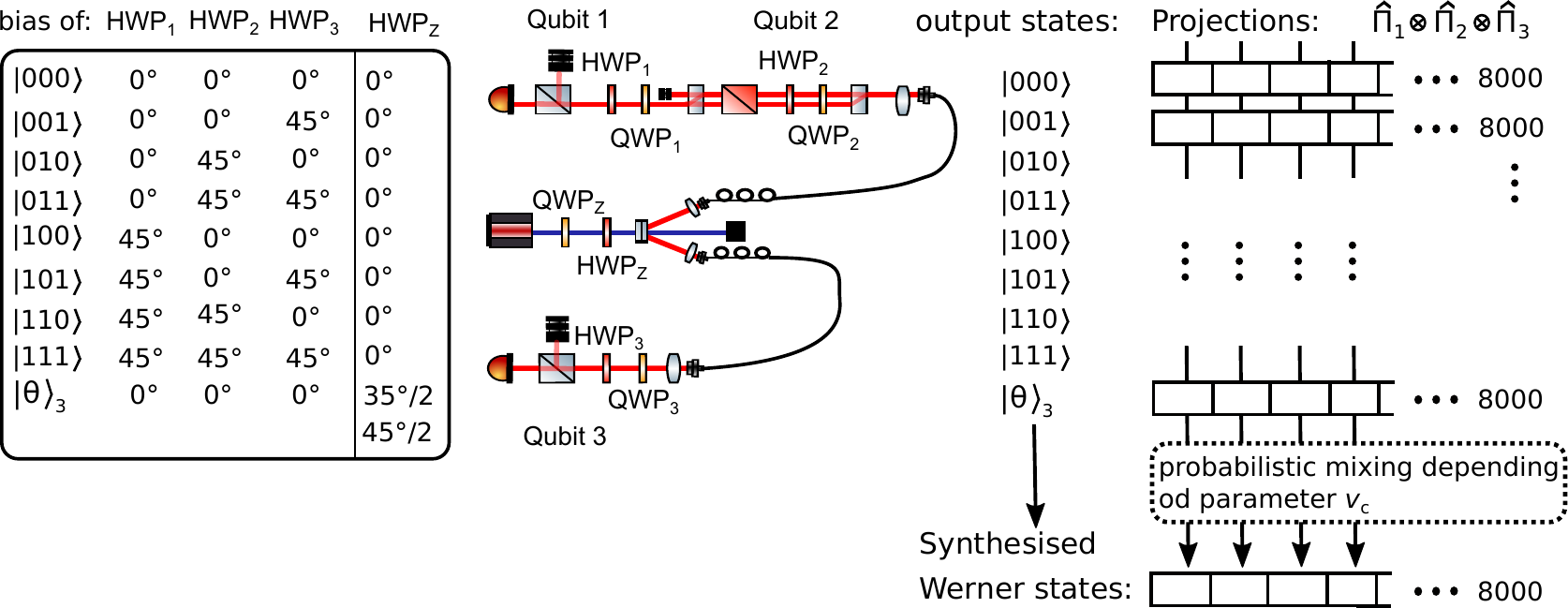}\hspace*{\fill}
\caption{Scheme of synthesising procedure for all $8 000$ random projections $\ket{\hat{\Pi}_1 \otimes \hat{\Pi}_2 \otimes \hat{\Pi}_3}$. In the left side of this scheme we provide HWP settings for each 3-qubit state. The final state $\hat{\rho}_{W}$ is mixed according to a prescription given in Eq.~\eqref{eq:Wmix}.}
\label{fig:Exschema}
\end{figure*}

The fact that $F(\rho^{\textrm{expt}}_{\theta}) < 1$ is naturally caused by the presence of experimental imperfections such as the improper setting of individual components or depolarization effects. Consequently, an effective form of the generated state should be considered as the three-qubit Werner-like state $\rho_3(\theta,v_{\theta})$ in Eq. \eqref{eq:rho3_theta}, where $v_{\theta}$ is associated with the strength of the effective noise inherently present during the experiment. 
The presence of such noise is certified by a reduction in purity, $P(\rho) = \textrm{Tr}(\rho^2)$, of the output state. By straightforward calculations we have found that $P(\rho^{\textrm{expt}}_{35^{\circ}})=0.982\pm0.005$ and $P(\rho^{\textrm{expt}}_{45^{\circ}})=0.976\pm0.005$. Then, using the relation $P(\rho) = \tfrac{1+7v_0^2}{8}$ \cite{BarasinskiPhysRevA2019} the visibility $v_0$ has been estimated. In our case the visibility is equal to $v_{35^{\circ}} = 0.990\pm0.003$ and $v_{45^{\circ}} = 0.986\pm0.003$. These values are further utilized to establish an appropriate reference point of theoretical predictions.

Next, using the reconstructed output-state, $\rho^{\textrm{expt}}_{\theta}$, and numerical procedure described in Sec. \ref{theory_nonlocal}, the nonlocal fraction has been evaluated
\begin{eqnarray}
p_{\textrm{V}}(\rho^{\textrm{expt}}_{45^{\circ}}) &=  9.018 \pm 0.926\% ,  \nonumber\\
p_{\textrm{V}}(\rho^{\textrm{expt}}_{35^{\circ}}) &=   8.237 \pm 0.676\% .
\label{eq:pV_rho}
\end{eqnarray}
For each state $10^8$ different settings have been numerically examined. Comparing these results with theoretical prediction, $p_{\textrm{V}}(45^{\circ},v_{45^{\circ}}) =  8.830 \%$ and  $p_{\textrm{V}}(35^{\circ},v_{35^{\circ}}) =  8.279 \%$, we see a very good agreement between both sets of outcomes.

\subsection{Nonlocal fraction measurements - reference frames independent approach}
\label{nonlocal_fraction_RFI}

In the second step, we relax the experimental requirements and consider the reference frame independent approach. In this case, all three qubits of the desired $\rho^{\textrm{expt}}_{\theta}$ state are subjected to randomly chosen local projections $\ket{\hat{\Pi}_1 \otimes \hat{\Pi}_2 \otimes \hat{\Pi}_3}$. The whole process includes $n=8000$ projection settings. For each adjustment of $\theta$ and $\ket{\hat{\Pi}_1 \otimes \hat{\Pi}_2 \otimes \hat{\Pi}_3}$, we measure coincidence detections (\textit{CC}) over approximately $20~\textrm{s}$ and we registered one value of \textit{CC} per projection. 
The values of \textit{CC} are used to determine all correlation coefficient (see \cite{BarasinskiPhysRevA2019,Barasinskipra101_2020}) and then, to test all $185$ Bell inequalities relevant for the genuine multipartite nonlocal correlations \cite{Bancalpra88_2013}. Note that in this test all possible relabeling of parties, inputs, and outputs has been taken into account. The value of Bell inequality was determined with precision $\pm0.015$. Dividing the number of projection setting which provide violation of local realism by the total number of setting $n$, the nonlocal fraction has been estimated. We obtained the following results:
\begin{eqnarray}
p_{\textrm{V}}^{\textrm{CC}}(45^{\circ}) &= 8.594 \pm 1.546 \%  \nonumber\\
p_{\textrm{V}}^{\textrm{CC}}(35^{\circ}) &= 8.722 \pm 1.174 \%. 
\label{eq:pV_RFI}
\end{eqnarray}

As we see in Fig. \ref{fig:comparison_pV}, our results in Eq. \eqref{eq:pV_RFI} match correctly to the attainable range of theoretical predictions if the precision of $v_0$ is included. Specifically, for the error bar of $v_0$ equaling to $\pm0.003$ one obtains $8.302 \% \leq p_{\textrm{V}}(45^{\circ},v_{45^{\circ}}) \leq 9.377 \%$ and 
$7.735 \% \leq p_{\textrm{V}}(35^{\circ},v_{35^{\circ}}) \leq 8.848 \%$. However, the values of $p_{\textrm{V}}^{\textrm{CC}}(\theta)$ slightly differ from $p_{\textrm{V}}(\rho^{\textrm{expt}}_{\theta})$ in Eq. \eqref{eq:pV_rho}. In particular, $p_{\textrm{V}}^{\textrm{CC}}(45^{\circ}) < p_{\textrm{V}}^{\textrm{CC}}(35^{\circ})$. 
To explain such difference, we emphasize that due to inherent experimental fluctuation the generated state slightly varies over the course of the entire data acquisition time (about two days). For that reason, one may expect some fluctuations of the inherent noise arising due to e.g. dephasing and depolarization, to name just a few. 

In order to verify this conclusion, the distribution of the strength of violation for random measurements has been analyzed. In other words, we simulate a robustness of the nonlocal fraction $p_{\textrm{V}}^{\textrm{CC}}$ using the accumulated data for random sampling. As we see in Fig. \ref{fig:Nonlocal_dist}, for both values of angle $\theta$ the simulated relationship between $p_{\textrm{V}}^{\textrm{CC}}$ and $\mathcal{I}_{\min}$ has a similar shape as its theoretical counterpart (see Appendix \ref{appendixB}). Furthermore, by fitting our experimental data with Eq. \eqref{eq:v_Werner3b} we have found the following results $\{\theta, v_{45^{\circ}}\}\approx \{44.7^{\circ},0.984\}$ and $\{\theta, v_{35^{\circ}}\}\approx \{36.0^{\circ},0.996\}$ what is in line with our previous observations.

\subsection{Genuine concurrence measure - reference frames independent approach}

\begin{figure}
\centering
\hfill\includegraphics[scale=1.]{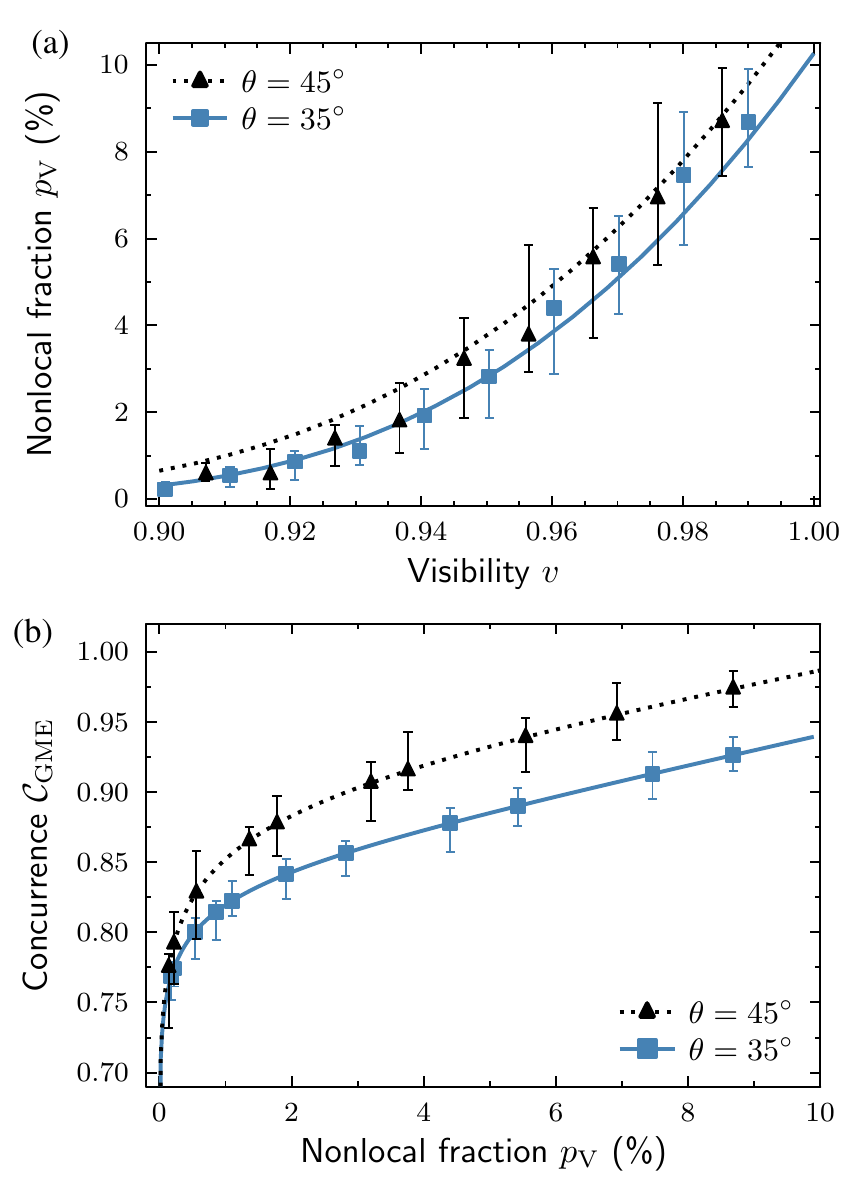}\hspace*{\fill}\\
\caption{(a) Dependence of nonlocal fraction on the visibility and (b) the relation between genuine concurrence and nonlocoal fraction for three-qubit Werner-like states. In both panels symbols denotes experimental measurements for $\theta = 35^{\circ}$ (triangles) and $\theta = 45^{\circ}$ (squares) while curves depict theoretical predictions.
}
\label{fig:pv35}
\end{figure}

The final stage of our experiment is to measure the GME-concurrence for the Werner-like states. In order to do that, the Werner-like states $\rho^{\textrm{expt}}_3(\theta,v)$ were synthesised with controlled visibility $v$ in range $\left[ 0.9; v_{\theta}\right]$. This is accomplished by controlled mixing (with probability $v_\textrm{c}$) the output state $\rho^{\textrm{expt}}_{\theta}$ and white noise, i.e. $v_\textrm{c} \rho^{\textrm{expt}}_{\theta}+(1-v_\textrm{c}) \rho_{\textrm{white noise}}$. As a result, one has 
\begin{eqnarray}
\rho^{\textrm{expt}}_3(\theta,v) &=& v_\textrm{c} v_{\theta} \,\ket{\theta}_3\bra{\theta}  + \nonumber \\
&+& \sum_{\textrm{basis}} \frac{1-v_\textrm{c} v_{\theta}}{8}\ket{\textrm{basis}}\bra{\textrm{basis}}, 
\label{eq:Wmix}
\end{eqnarray}
where the total visibility $v \equiv v_\textrm{c} v_{\theta}$ with $v_{\theta}$ being a constant value defined above and controlled parameter $v_\textrm{c}$ varying with a step $\delta_{\textrm{v}} = 0.01$. 

Now, to synthesise projection results $CC$ for any mixed state in Eq. \eqref{eq:Wmix}, the experimental setup was set to gradually generate 8 basis states $\ket{\textrm{basis}}$. Similarly as in the Sec. \ref{nonlocal_fraction_RFI}, each of the states was subjected to the same set of random projections $\ket{\hat{\Pi}_1 \otimes \hat{\Pi}_2 \otimes \hat{\Pi}_3}$ as these of $\rho^{\textrm{expt}}$ (including tomography projections). Finally, values of $CC$ were probabilistically mixed according to the following routine (Fig.~\ref{fig:Exschema})
\begin{equation}
CC_{i}(\theta,v) =  v_\textrm{c} \,CC_{i}(\rho^{\textrm{expt}}_{\theta}) + \sum_{\textrm{basis}}\frac{1-v_\textrm{c}}{8} CC_{i}(\ket{\textrm{basis}}), \label{eq:WCC}
\end{equation}
where $CC_{i}(\theta,v) \equiv CC_{i}(\rho^{\textrm{expt}}_3(\theta,v))$ and $CC_{i}(\rho)$ denotes the values of $CC$ for the state $\rho$ and the $i$-th projector $\ket{\hat{\Pi}^{(i)}_1 \otimes \hat{\Pi}^{(i)}_2 \otimes \hat{\Pi}^{(i)}_3}$. This procedure results in $8 000$ values of CC for each generated state $\rho^{\textrm{expt}}_3(\theta,v)$ that can be further analysed. Note that for every state, all $CC_{i}  $ were normalised with respect to the overall generation rate for the particular state.

On the basis of these results, the nonlocal fraction of $\rho^{\textrm{expt}}_3(\theta,v)$ has been determined. As we see in Fig.~\ref{fig:pv35}(a), our measurements are in good agreement with theoretical predictions given in Eq. \eqref{eq:v_Werner3}. Finally, using Eq. \eqref{eq:C_Werner3b} the GME-concurrence for the Werner-like states has been establishes and the accomplished results are in perfect agreement with theory Fig.~\ref{fig:pv35}(b).

\section{Conclusions}
\label{ch:concl}

In conclusion, we have theoretically and experimentally investigated the entanglement-assessment protocol for two- and three-qubit Werner-like states. Our proposal is based on the concept of the nonlocal fraction which denotes the probability of detection of nonlocal correlation under randomly measurements. 
Using numerical calculations we have found the relationship between the degree of entanglement and nonlocal fraction. Then, our method has been successfully applied to the experimental measurements of the GME-concurrence of the three-qubit Werner-like state, revealing perfect agreement with theoretical predictions. 

The advantage of using random sampling in our protocol is a great simplification of experimental procedures as the alignment and calibration of remote devices are not necessary anymore. Therefore, our protocol can be applied in an unstable environment, where the above-mentioned requirements are hard to be met. 

Although in this paper we focus on the Werner-like states, our protocol can also be used for an arbitrary mixed state. In this broader context, the protocol can operate as an indicator of a lower bound of entanglement for the state under consideration. From the point of view of quantum communication, such finding is of great importance as it allows to characterize a minimal efficiency on the communication protocol.

\begin{acknowledgments}
The authors thank Cesnet for providing data management services. Numerical calculations were performed in the Wroclaw Centre for Networking and Supercomputing, Poland. K.J., A.B, A.Č., and K.L. acknowledge financial support by the Czech Science Foundation under the project No. 20-17765S. The authors also acknowledge project CZ.02.1.01/0.0/0.0/16\_019/0000754 of the Ministry of Education, Youth and Sports of the Czech Republic. K.J. also acknowledges the Palack\'y University internal grant No. IGA-PrF-2021-004.

\end{acknowledgments}

\appendix
 
\section{Analytical Derivation of Eq. \eqref{eq:pV_Werner2}}
\label{appendixA}

The CHSH inequality for general two-qubit state $\rho$ can be written as \cite{Horodeckipla200_1995}
\begin{eqnarray}
|\vec{a}_0 \cdot R^{\rho} \cdot \left( \vec{b}_0+\vec{b}_1\right) + \vec{a}_1 \cdot R^{\rho} \cdot \left( \vec{b}_0-\vec{b}_1\right)| \leq 2,
\label{eq:a1}
\end{eqnarray}
where $\vec{a}_0$, $\vec{a}_1$, $\vec{b}_0$, $\vec{b}_1$ are unitary vectors in $\mathbb{R}^3$ and $R^{\rho}$ denotes the $3 \times 3$ correlation matrix with elements $R^{\rho}_{ij} = \textrm{Tr} [\rho \left(\sigma_i \otimes \sigma_j\right)]$ given in terms of the three Pauli matrices. For the special case, when $\rho$ stands for the Werner state (in the form proposed in Ref. \cite{Wernerpra40_1989}) the correlation matrix $R=-v \id_3$, where $v$ is the visibility.

Next we introduce a pair of unitary vectors $\vec{c}_0$ and $\vec{c}_1$ by
$\vec{b}_0+\vec{b}_1 = \vec{c}_0 \sqrt{2 (1+x)}$, 
$\vec{b}_0-\vec{b}_1 = \vec{c}_1 \sqrt{2 (1-x)}$,
where $x=\vec{b}_0 \cdot \vec{b}_1$. Substituting all these quantities into Eq. \eqref{eq:a1} one has 
\begin{eqnarray}
|\vec{a}_0 \cdot \vec{c}_0 \sqrt{1+x} + \vec{a}_1 \cdot \vec{c}_1 \sqrt{1-x}| \leq \frac{\sqrt{2}}{v}.
\label{eq:a2}
\end{eqnarray}

To prove Eq. \eqref{eq:pV_Werner2} we shell find how often inequality \eqref{eq:a2} is violated when unit vectors $\vec{a}_0$, $\vec{a}_1$, $\vec{c}_0$, $\vec{c}_1$ and the variable $x$ are chosen independently, randomly, and isotropically. Following arguments presented in Ref. \cite{Liangprl104_2010}, to solve the above problem, it is sufficient to sample $x$ and dot products $\vec{a}_0 \cdot \vec{c}_0$ and $\vec{a}_1 \cdot \vec{c}_1$  uniformly from the interval $[-1,1]$ as the actual direction of individual vectors is irrelevant (hereafter, we use $\alpha = \vec{a}_0 \cdot \vec{c}_0$ and $\beta = \vec{a}_1 \cdot \vec{c}_1$).

From a geometrical point of view, this solution denotes the fraction of the cube's volume containing points $(\alpha, \beta,x)$ violating the inequality \eqref{eq:a2}. For particular fixed $x$ the regime of the cube containing points violating Eq. \eqref{eq:a2} are given by 
\begin{eqnarray}
\beta &>& \frac{\sqrt{2} - \alpha v \sqrt{1+x}}{v \sqrt{1-x}},\nonumber\\
\beta &<& -\frac{\sqrt{2} + \alpha v \sqrt{1+x}}{v \sqrt{1-x}}.
\label{eq:a3}
\end{eqnarray}
Therefore, with some straightforward calculation, one can find that the fraction of Alice and Bob's measurement directions that would violate the CHSH inequality and hence, the nonlocal fraction is given by
\begin{eqnarray}
p_{\textrm{V}} = 4 \int_{x_{-}}^{x_+} \frac{\left(\sqrt{2} - v (\sqrt{1 - x} + \sqrt{1 + x})\right)^2}{V_{\textrm{cube}}~v^2 \sqrt{1 - x^2}} \,dx,
\label{eq:a4}
\end{eqnarray}
where $V_{\textrm{cube}}=2^3$ stand for the cube's volume, the integration is performed for $x_{\pm} = \pm \sqrt{\frac{2 v^2-1}{v^4}}$ and the result is multiply by $4$ to take this into account any possible relabeling of measurement settings and/or outcomes. It is because, for any given measurement directions at most one of CHSH  inequalities can be violated. The value of $x_{\pm}$ is caused by the fact that for fixed $v$ and $x>x_{+}$ ($x<x_{-}$) there is no pairs $(\alpha,\beta)$ (both in the interval $[-1,1]$) which satisfy constraints \eqref{eq:a2}. After appropriate integration in Eq. \eqref{eq:a3}, we get Eq. \eqref{eq:pV_Werner2}. Note that for $v=1$ the nonlocal fraction $p_V=2 (\pi-3)$ what is in line with \cite{Liangprl104_2010}.

\section{Nonlocal fraction based on the distribution of the strength of violation}
\label{appendixB}

Let us take a three-qubit state $\rho$ and a finit set of measurement settings $\{\hat{M}_i\}$, where $i=1, \dots, m$. To verify whether the genuine nonlocal correlations are generated for the state $\rho$ and given measurement setting $\hat{M}_i$, one should test $185$ Bell inequalities \cite{Bancalpra88_2013} of the form $\tilde{\mathcal{I}}_j(\rho| M_i) \leq C^{\textrm{LHV}}_j$, where $j=1, \dots, 185$. To this end, it is expedient to consider $C^{\textrm{LHV}}_j=1$ and $\mathcal{I}_j(\rho| M_i) = \tilde{\mathcal{I}}_j(\rho| M_i)/C_{\textrm{LHV}}$. Based on such test, a maximal strength of violation for $\hat{M}_i$ is determined as $\mathcal{I}^{\max}_i(\rho) = \max\limits_j\{\mathcal{I}_j(\rho| M_i)\}$, where the maximum is taken over $185$ Bell inequalities. Dividing the number of $\mathcal{I}^{\max}_i(\rho)$, satisfying the constraints $\mathcal{I}^{\max}_i(\rho)>1$, by the number of measurement settings $m$, the nonlocal fraction is estimated 
\begin{equation}
p_{\textrm{V}}(\rho)= \lim_{m \to \infty} \frac{n \Big(\{\mathcal{I}^{\max}_i(\rho), \mathcal{I}^{\max}_i(\rho)>1\} \Big)}{m}.
\label{eq:b1}
\end{equation}

Next, let us consider a state $\sigma(v) = v \rho + (1-v)/8 \cdot \id_8$, i.e. a statistical mixture of the state $\rho$ and white noise. Then, one can easily prove that $\mathcal{I}_j(\sigma| M_i) = v~\mathcal{I}_j(\rho| M_i)$ and hence, the maximal strength of violation $\mathcal{I}^{\max}_i(\sigma) = v~\mathcal{I}^{\max}_i(\rho)$. Consequently, by analogy to Eq. \eqref{eq:b1}, the nonlocal fraction of state $\sigma$ can be written 
\begin{equation}
p_{\textrm{V}}(\sigma)= \lim_{m \to \infty} \frac{n \Big(\{\mathcal{I}^{\max}_i(\rho), \mathcal{I}^{\max}_i(\rho)>\mathcal{I}_{\min}=1/v\} \Big)}{m}.
\end{equation}
In other words, if one knows the distribution of the strength of violation $\{\mathcal{I}^{\max}_i(\rho)\}$, then the nonlocal fraction of any state $\sigma(v)$ can be estimated by suitable shiftiness of the classical threshold denoted by $\mathcal{I}_{\min}$. As a result, one can find a relationship between $p_{\textrm{V}}(\sigma)$ and the visibility $v$ (c.f. Fig. \ref{fig:Pure3}(a)).

In particular, if we assume that $\rho = \rho_3(\theta,v_0)$, then the state $\sigma(v)= v\cdot v_0 \ket{\theta}_3\bra{\theta} + \frac{1-v\cdot v_0}{8} \id_8$ and the relationship between $p_{\textrm{V}}(\sigma)$ and $v$ is described by Eq. \eqref{eq:v_Werner3} with unknown values $v_0$ and angle $\theta$. Therefore, Eq. \eqref{eq:v_Werner3} can be rewritten as
\begin{eqnarray}
v=\frac{1}{v_0} \left(v_3^{\textrm{cr}}(\theta)+g_1(\theta)~p_{\textrm{V}}^{1/6} + g_2(\theta)~p_{\textrm{V}}^{1/2} +g_3(\theta) ~p_{\textrm{V}}\right).\nonumber\\
\label{eq:v_Werner3b}
\end{eqnarray}
By fitting the distribution $p_{\textrm{V}}(\sigma)$ vs. $v$ described above with Eq. \eqref{eq:v_Werner3b} one gets an approximation of both parameters $v_0$ and $\theta$.

\end{document}